\documentclass{article}
\usepackage{iclr2027_conference,times}

\usepackage{amsmath,amsfonts,bm}

\def\eqref#1{equation~\ref{#1}}

\def\1{\bm{1}}

\DeclareMathAlphabet{\mathsfit}{\encodingdefault}{\sfdefault}{m}{sl}
\SetMathAlphabet{\mathsfit}{bold}{\encodingdefault}{\sfdefault}{bx}{n}

\usepackage{amsmath,amssymb}
\usepackage{booktabs}
\usepackage{array}
\usepackage{graphicx}
\usepackage{hyperref}
\usepackage{url}
\hypersetup{pdfauthor={Suxin Ji, Hungtao Wan, Shaoxuan Chen, An Zhang}}

\graphicspath{{figures/}{results/}}

\title{Semantic Behavioral Watermarking:\\
Paraphrase-Robust and Forgery-Resistant\\
Provenance for LLM Agents}

\author{Suxin Ji$^{1}$\quad Hungtao Wan$^{1}$\quad Shaoxuan Chen$^{2}$\quad An Zhang$^{3}$ \\
$^{1}$Independent Researcher \\
$^{2}$University of Massachusetts Amherst\quad $^{3}$Chongqing University}
\iclrfinalcopy

\begin{document}
\maketitle
\lhead{}\renewcommand{\headrulewidth}{0pt}
\begin{abstract}
Behavioral watermarking embeds an owner identifier in an LLM agent's high-level action choices,
giving provenance without touching output tokens. Prior agent watermarks bind the signal to raw
strings and break in two ways. AgentMark's own robustness test paraphrases the \emph{observation}
alone and bit-recovery falls to 16.8\%; its released script holds the per-step key fixed, so that
experiment measures distribution drift, not the key desynchronization its context-hash design
separately admits. All three prior schemes also bind the watermark to the \emph{exact action
symbol}, so renaming a tool desynchronizes decoding even when the observation is untouched --- the
axis we measure and close. And every prior agent watermark studies only \emph{removal}: none asks
whether an adversary can forge a trajectory that verifies as someone else's, a question answered
affirmatively for text watermarks \citep{jovanovic2024stealing}. We present Semantic Behavioral
Watermarking (SBW): watermarking over \emph{semantic action clusters} under history conditioning,
with the public-cluster bin replaced by keyed collision-resistant binning whose fresh-bucket
assignment is provably unpredictable in the random-oracle model. Across five agent models
(3B--14B, four vendors) and three encoders the ordering holds on both benchmarks: on ToolBench
(600 trajectories per model) detection under rewriting is 0.49--0.66 for cluster-level versus
0.05--0.17 for exact-symbol at a permutation-calibrated 1\% FPR, at 72--83\% choice agreement
against 22--27\% for logit biasing; on ALFWorld (100 episodes per model) it is 0.92--0.97 versus
0.00--0.01. Keyed binning takes adaptive forgery from 100\% to the false-positive floor at the
primary operating point (bge, r=64). We also mark the boundary that guarantee does \emph{not}
cover: when the adversary copies the victim's own steps, shuffled splicing is neutralized (0.000 on
Qwen2.5-3B) but chained replay remains at 0.76--0.98 across the five models, reported as open. Paraphrase robustness costs about half
of the per-step watermark capacity.
Code is available at \url{https://anonymous.4open.science/r/SBW-Agent-Watermark}.
\end{abstract}

\section{Introduction}
\begin{figure}[t]\centering
\includegraphics[width=\linewidth]{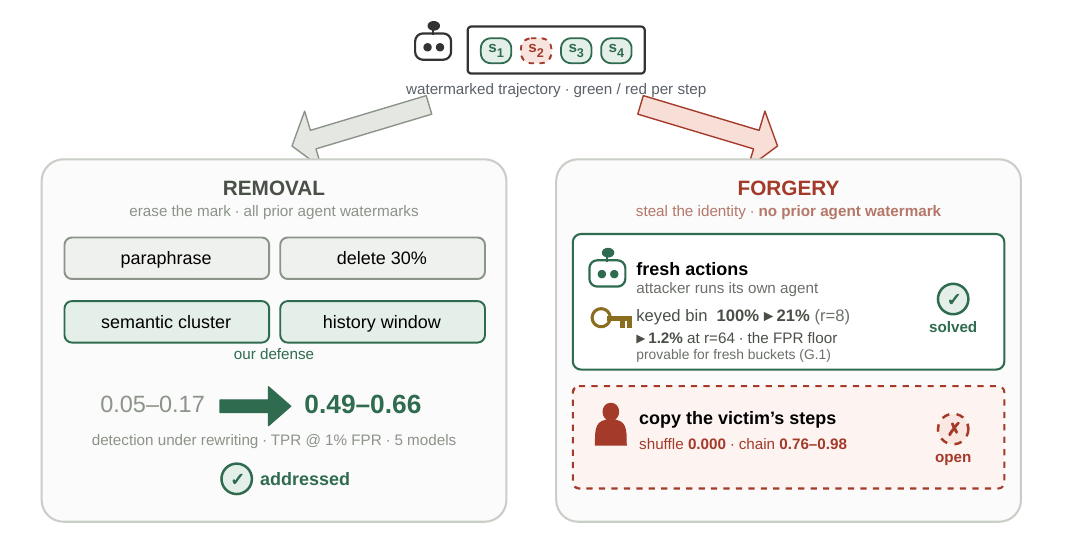}
\caption{\label{fig:overview}\textbf{The two attack directions, and which one prior work leaves open.} \textbf{Left --- removal:} the adversary wants the owner's mark to stop being detectable. Every prior agent watermark studies only this side and keys on the exact action symbol, so renaming a tool desynchronizes decoding; watermarking the \emph{semantic cluster} under history conditioning is what lifts detection under rewriting from 0.05--0.17 to 0.49--0.66 across five models. \textbf{Right --- forgery:} the adversary wants \emph{its} trajectory to verify as the owner's. We split this side explicitly: against \emph{fresh} actions keyed binning drives forgery from 100\% to 21\% at the default r=8 and to 1.2\% at r=64 --- the false-positive floor (provable for fresh buckets, Thm.~G.1), but when the adversary \emph{copies} the victim's steps nothing must be predicted and the key does not help --- shuffled splicing is neutralized (0.000, measured on Qwen2.5-3B) while chained replay stays at 0.76--0.98 across the five models (Table~\ref{tab:copystyle}), which we report as an open boundary rather than close.}
\end{figure}

LLM agents that plan and call tools are increasingly deployed as commercial assets; owners need to prove that a given trajectory was produced by their agent (provenance) and to deter unauthorized reuse (IP protection). Token-level text watermarks do not apply: the ownership signal lives in the agent's discrete behavior --- which tool or subgoal it selects --- not in surface tokens. A recent line of behavioral watermarks embeds the signal there: Agent Guide \citep{huang2025agentguide} biases behavior probabilities; AgentMark \citep{huang2026agentmark} embeds multi-bit identifiers via distribution-preserving conditional sampling; SeqWM \citep{an2026seqwm} conditions the green/red split on the recent action history for position-agnostic, deletion-robust verification.
Two gaps remain. (G1) Paraphrase fragility. All three key the watermark on the exact action symbol. A semantics-preserving rewrite changes that symbol and desynchronizes decoding, and SeqWM's history windows hash exact strings. That binding to raw strings is demonstrably fragile: AgentMark's own robustness test paraphrases the \emph{observation} block --- its released script preserves the action list verbatim and teacher-forces the same per-step key for both runs --- and bit-recovery falls to 16.8\% at a mean KL of 3.23, i.e. the measured failure is drift in the elicited distribution. Its per-step key is separately a hash of the preceding context, so key desynchronization is a second exposure that this experiment does not measure. Action renaming is the orthogonal axis, the one an exact-symbol design leaves exposed, and the one we measure directly (EXACT 0.05--0.17 under rewriting, \S 6.1). (G2) Forgery. Every agent watermark to date studies only removal robustness. But behavioral watermarks that use a public semantic encoder introduce a new attack surface: an adversary who observes a few watermarked trajectories can estimate the green/red structure and forge a trajectory that verifies as another owner --- enabling framing or repudiation.
We address both with Semantic Behavioral Watermarking (SBW), sketched in Fig.~\ref{fig:overview}: (1) cluster-level history-conditioned watermarking (\S 4.1) --- a SeqWM-style scheme over semantic-cluster symbols, so both the current symbol and the history window survive rewriting; (2) keyed collision-resistant binning (\S 4.2) --- bin(a) = H(sk, $quantize(R_{sk}\cdot$ E(a))), provably unpredictable on fresh buckets (Thm. G.1) and reducing real forgery from 100\% to $\sim$21\%; (3) a synchronization-stability and elicited-capacity theory (\S 5) with proven directional margins; (4) a real-data evaluation (\S 6; Fig.~\ref{fig:results}) with multi-seed, FPR-calibrated detection, matched-null controls, an adversarial verification, and native head-to-heads against both a released implementation (AgentMark) and a from-paper reimplementation (AGENTWM).
To our knowledge this is the first behavioral watermark for agents that is both paraphrase-robust and forgery-resistant. Beyond (1)--(4) it contributes a previously-unstudied forgery attack on semantic behavioral watermarks, a provable fresh-bucket guarantee with a precisely characterized residual, and a quantified boundary for the copy-style attacks that guarantee does \emph{not} cover.

\section{Related Work}
Token and semantic text watermarking. KGW \citep{kirchenbauer2023watermark} partitions the vocabulary into a context-seeded green list; paraphrase degrades it measurably \citep{kirchenbauer2024reliability}, and distortion-free constructions \citep{kuditipudi2024distortionfree} address the utility side of the same trade. The semantic-text line fixes this by clustering the embedding space: SemStamp \citep{hou2024semstamp} (LSH partitions), k-SemStamp \citep{hou2024ksemstamp} (k-means partitions), SIR \citep{liu2024semanticinvariant} (semantics-conditioned logit bias), and PASA \citep{ai2026pasa} (embedding-space watermark with detection-optimality theory). Clustering the tool space is itself not new --- Tool-Planner \citep{liu2025toolplanner} groups tools by k-means to organize task planning --- but it has been used for planning efficiency, never as the unit that carries an ownership signal. SBW ports the cluster-as-watermark-unit idea to the multi-step agent-action setting, where the watermarked unit is a discrete tool choice, robustness must hold across held-out tools, and there is no free rejection-sampling knob.
Agent behavioral watermarking. Agent Guide $\rightarrow$  AgentMark $\rightarrow$  SeqWM are the direct lineage; AGENTWM \citep{wang2026agentwm} biases the distribution over functionally-equivalent tool paths for IP; ActHook \citep{meng2026acthook} inserts trigger actions into training trajectories; MemMark \citep{zhang2026memmark} watermarks memory writes; and \citet{nian2026implicit} do implicit execution tracing for multi-agent attribution. Table~\ref{tab:related} summarizes how SBW differs from each.
Watermark stealing and spoofing. For \emph{text} watermarks the forgery direction is established: \citet{jovanovic2024stealing} show that an adversary who queries a watermarked model can approximate its green-list rule well enough to spoof it. No agent behavioral watermark has been examined this way --- the agent setting substitutes a public semantic encoder for the vocabulary partition, which is what makes the structure recoverable from observation alone, and what the keyed binning of \S 4.2 is designed to remove.
\begin{table}[t]
\caption{how SBW differs from the closest prior work.}
\label{tab:related}
\begin{center}\small\setlength{\tabcolsep}{4pt}
\begin{tabular}{@{}>{\raggedright\arraybackslash}p{0.135\linewidth}>{\raggedright\arraybackslash}p{0.385\linewidth}>{\raggedright\arraybackslash}p{0.44\linewidth}@{}}
\toprule
vs. & What it does & What SBW adds \\
\midrule
AgentMark & distribution-preserving sampling over exact tool choices & watermarks over semantic clusters $\rightarrow$  paraphrased actions decode to the same bin \\
SeqWM & history-conditioning on exact action strings & history over cluster symbols $\rightarrow$  robust to paraphrase and deletion \\
AGENTWM & biases enumerated equivalent tool paths (closed set) & learned, open-vocabulary clusters $\rightarrow$  generalizes to held-out tools \\
k-SemStamp, PASA, SIR & semantic watermarking of single-stream text & multi-step agent actions; keyed anti-forgery; capacity theory \\
all above & study only removal robustness & first forgery study for agent watermarks + keyed defense \\
\bottomrule
\end{tabular}
\end{center}
\end{table}
\section{Threat Model}
An owner deploys a watermarked agent with secret key sk, a frozen public semantic encoder E, and a cluster assignment C. Verification observes a (possibly attacked) action trajectory. The owner controls generation-time decoding and elicits an explicit behavior distribution (as in AgentMark); capacity is measured only under this deployed elicitation operator $\Phi$  and decoding config $\theta(\S$ 5.2), making elicitation part of the model rather than a hidden variable.
Robustness attacks (in scope): semantics-preserving rewriting of actions/observations (including renaming to unseen tools) and step deletion/truncation. Cross-type policy flips $(explore\rightarrow$ grab), which change task semantics, are out of scope by construction and are not claimed to be detected. Forgery attack (in scope, new): an adversary holding the public encoder E and M watermarked trajectories produces a new trajectory over \emph{fresh} actions that passes the owner's detector without sk. \textbf{Fresh-action scope.} The restriction is deliberate: a replayed trajectory \emph{is} the victim's watermarked output, so detecting it is attribution working, not a forgery (the copy-paste analogue in text watermarking); an attacker running their own agent on their own tasks necessarily acts with fresh actions; and Thm. G.1 covers exactly the fresh-bucket branch. Replay, splicing and chaining are quantified separately in \S 6.2 and are \emph{not} claimed to be solved.
\section{Method}
SBW is entirely training-free: a frozen off-the-shelf sentence encoder E, one k-means quantization of the action-embedding space, a keyed pseudo-random function, and a permutation-test detector. No gradient step, no fine-tuning, no learned watermark parameters --- swapping the agent or the encoder is pure inference, and the anti-forgery guarantee is a property of the keyed construction rather than of any trained artifact.
\subsection{Cluster-level history-conditioned watermark}
Let $s_{t}$ = $C(a_{t})$ be the semantic-cluster symbol of the action at step t and $c_{t}$ = $(s_{t-w},\ldots,s_{t-1})$ the recent-cluster history. A step is green iff a keyed PRF PRF(sk, $c_{t},$ $s_{t})$ returns 1. During generation, with probability $p_{green}$=0.95 (the detection and copy-style operating point; \S 6.2 states the forgery study's own) the step is resampled from the green candidates in proportion to their elicited mass, so the bias is support-restricted (no probability is moved onto actions $P_{t}$ did not already admit) but not distribution-preserving in AgentMark's sense (\S 6.1 measures the cost). Otherwise, and whenever no candidate is green, the step falls back to the unwatermarked argmax and carries no bit. Detection uses SeqWM's random-key permutation test: score = number of green windows over the observed cluster-symbol stream; p-value from M wrong keys. Because a paraphrase of any action maps to the same cluster, both $s_{t}$ and the history $c_{t}$ survive rewriting; deletion is handled by the position-agnostic window score exactly as in SeqWM. The history length w is a \emph{deployment knob}, not a free parameter: a sweep $w\in$ \{1,2,3\} shows longer contexts raise clean detection (0.83/0.91/0.98) but degrade deletion robustness sharply (CLUSTER TPR at 30\% deletion 0.48/0.31/0.24), since each deleted step destroys $w{+}1$ windows, while immunity to step-shuffled recombination is w-independent ($\approx$ 0 at every w). We use w=1 throughout unless stated, i.e. the deletion-robust end of that trade.
\subsection{Keyed collision-resistant binning (anti-forgery)}
A naive scheme sets green over public clusters, so an adversary with E can recover which clusters are green and steer fresh actions into them (\S 6.2). We instead key the binning: bin(a) = H(sk, $quantize(R_{sk}$ $\cdot$  E(a))), where $R_{sk}$ is a secret random projection to r hyperplanes. The bins are now orthogonal to the public embedding geometry. Theorem G.1 shows that for a fresh action whose keyed bucket has not been observed, the adversary's bin-prediction advantage is negligible in the random-oracle model; in the observed-bucket regime (empirically $\sim$98\% of fresh actions at r=8; \texttt{bucket\_collision.json}, whose generating script was not retained), the keyed projection blunts the nearest-neighbor green-bit transfer that public-cluster binning admits --- reducing real forgery from 100\% to $\sim$21\% (\S 6.2).
\subsection{Margin/abstain code}
Boundary actions (small cluster margin) are the source of decode errors. SBW routes actions with margin below $\varepsilon$ * to an abstain state (they carry no bit but preserve synchronization), trading a measured amount of capacity for near-zero error on the carried bits.
\section{Theory}
\subsection{Synchronization stability (paraphrase robustness)}
Define the margin $\gamma(a)$ = $1/2(d(E(a),\mu_{2})$ - $d(E(a),\mu_{1})),$ the half-gap to the second-nearest centroid.
\begin{itemize}
\item Prop. E.1 (bin invariance). If $\Vert$ $\Vert E(a')-E(a)\Vert\leq\varepsilon<\gamma(a)$, then bin(a') = bin(a) (Voronoi-cell stability).
\item Prop. E.2 (bit-error bound). Under displacement $\leq\varepsilon,$ the per-step flip rate $\rho(\varepsilon)$ $\leq$  $Pr_{a}[\gamma(a)$ $\leq\varepsilon].$
\item Prop. E.3 (detection). With per-step green bias $\delta,$ the permutation z-score scales as $\sqrt{T}\cdot(1-2\rho)\cdot\delta;$ detection is recoverable for any $\rho<1/2$ given enough steps T.
\item Cor. E.4 (margin/abstain). Abstaining on $\gamma\leq\varepsilon$ \textit{ drives $\rho\rightarrow$ 0 on carried bits at capacity cost $Pr[\gamma\leq\varepsilon$ }].
\item Prop. E.1'/E.2' (directional tightening, proven). The norm-based bounds are worst-case and loose $(\delta\approx$ 0.43 $\gg\gamma\approx$ 0.059). With the oriented margin $\gamma\perp(a)$ = $(\Vert$ $E(a)-\mu_{2}\Vert^{2}-\Vert$ $E(a)-\mu_{1}\Vert^{2})/(2\Vert\mu_{2}-\mu_{1}\Vert)$ and the dangerous projection $\Delta\perp(a)$ = $\langle\Delta(a),$ $u_{2}\rangle,$ a bisector lemma (proven both ways; residual $5.1{\times}10^{-16}$ on real tools) gives $\Delta\perp>\gamma\perp\Longleftrightarrow$  the rewrite is closer to $\mu_{2}$ than $\mu_{1},$ hence $\rho_{c_2}\leq Pr[\Delta\perp>\gamma\perp]\leq\rho$. On 2128 real tools: $\gamma\perp$  mean 0.1482 vs $\gamma$  0.0586; $\Delta\perp$  mean 0.0207 vs $\delta$  0.429 $(\geq$ 95\% of rewrite displacement is tangent to the dangerous direction); $Pr[\Delta\perp\geq\gamma\perp]$ = 0.0841, a ~$12\times$  tightening of the trivial 0.99 bound; the sandwich $\rho_{c_{2}}$ = 0.0606 $\leq$  0.0841 $\leq\rho$  = 0.1170 holds exactly; the safe condition gives flip-to-$c_2$ = 0.0000.
\end{itemize}
\subsection{Elicited watermark capacity}
The per-step channel is defined by the elicited distribution $P_{t}^{\Phi,\theta}:$ capacity $C(\Phi,\theta)$ = $E_{t}[cap(P_{t}^{\Phi,\theta})]$ is a joint function of policy and elicitation. Prop. F.1: C $\leq$  $E_{t}[H(P_{t})],$ vanishing for deterministic policies. Prop. F.2: cluster capacity $\leq$  tool capacity, gap shrinking in k. Prop. F.3 (payload budget): recovering an L-bit ID at efficiency $\eta$  under per-step error $\rho$  needs T $\geq$  $L/(\eta\cdot$ $C\cdot(1-h(\rho)))$ steps; calibrated (C=0.632, $\rho$ =0.117): L=128 $\rightarrow$  T $\geq$  422 steps --- motivating per-corpus aggregation for short episodes (30 documents of G=20 trajectories each). Our elicitation study makes F concrete: two elicitation operators (option-logprob vs explicit-ask) yield per-step entropies correlating at only 0.13 (\texttt{m2\_capacity.py}; the one main-text number that needs a GPU rollout rather than the frozen artifacts) --- capacity is undefined without fixing $\Phi$ .
\section{Experiments}
Setup. Real ToolBench \citep{qin2024toolllm}: 600 real tasks, 100 from each of the six G1/G2/G3 sub-splits, one trajectory per task; a local agent runs a bounded ReAct loop --- Qwen2.5-3B-Instruct and LLaMA-3.2-3B first, then Qwen2.5-7B, Mistral-7B and phi-4 (14B) under the same frozen harness (\S 6.1), eliciting an explicit $action_{weights}$ distribution at each step $\rightarrow$  600 real trajectories per model (30 documents of G=20; a tolerant parser keeps uniform-fallback below 4.4\% on every model). Clusters are k-means over bge-base embeddings of the 2128 unique real tools; paraphrase attacks rewrite each tool's name+description with a separate model. Detection is the random-key permutation test (M=300); all detection numbers are TPR at a permutation-calibrated 1\% FPR, $mean\pm$ sd over 5 seeds. An early 120-trajectory pilot showed the same ordering (App. A); all main-text numbers are the 600-trajectory runs.
\begin{figure}[t]\centering\includegraphics[width=\linewidth]{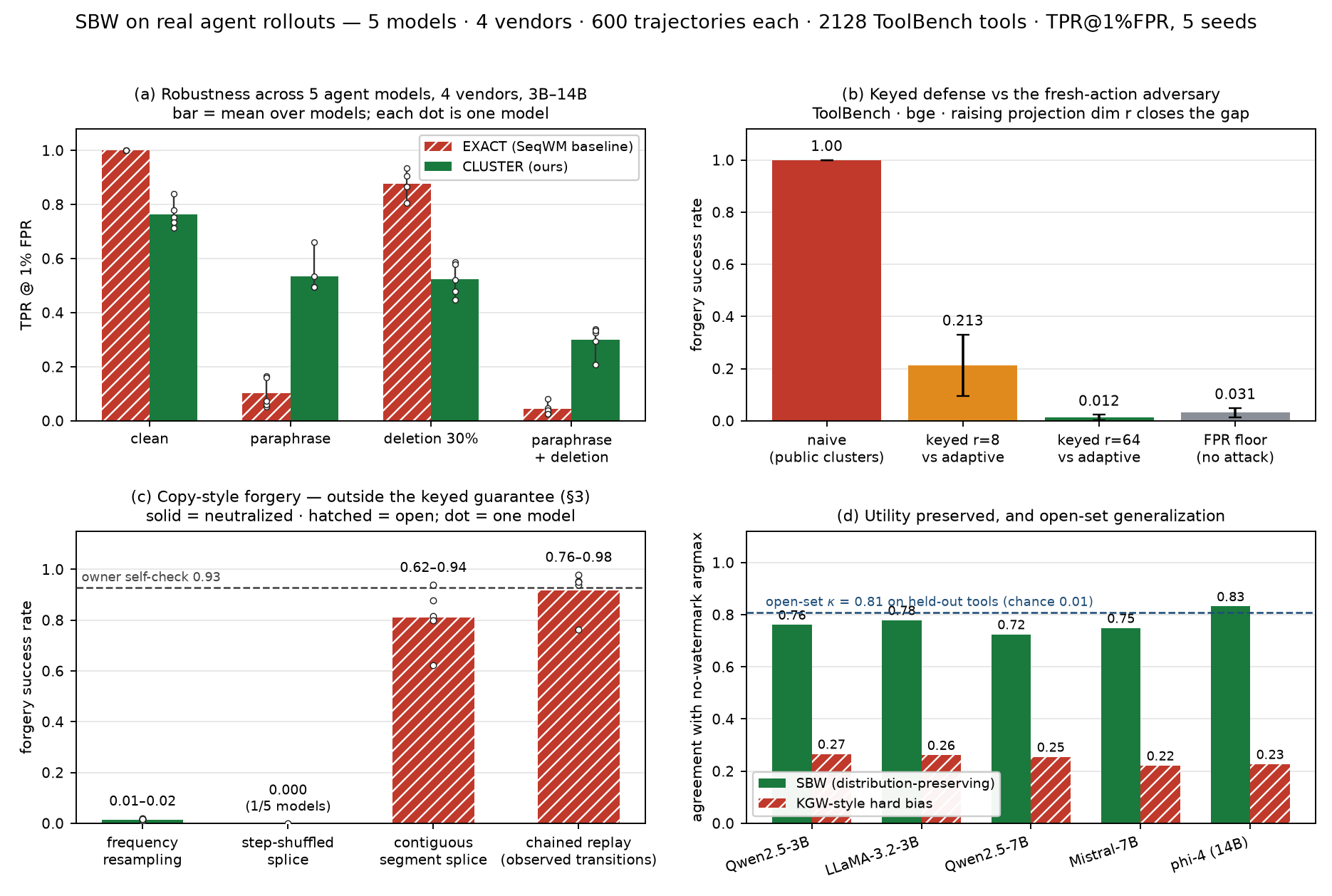}\caption{\label{fig:results}SBW on real agent rollouts, 600 trajectories per model. \textbf{(a)} The paraphrase-robust ordering across five agent models from four vendors; each dot is one model, so the spread \emph{is} the cross-model range. \textbf{(b)} Against the fresh-action adversary, raising the keyed projection dimension takes forgery to the permutation-FPR floor. \textbf{(c)} The copy-style attacks outside that guarantee (\S 3): recombination is neutralized, but contiguous splicing and chained replay approach the owner's own detection rate. \textbf{(d)} Utility and open-set generalization. In (c) solid bars are neutralized attacks and hatched bars open ones, so the panel reads in black and white. Regenerated from the released artifacts by \texttt{sbw/real\_plots\_600.py} (CPU-only).}\end{figure}
\subsection{Headline: the $2\times$ 2 under rewriting, cross-model, FPR-audited}
Table~\ref{tab:main2x2} reports the headline result; Figure~\ref{fig:results} gives it together with the forgery, copy-style and utility results at a glance. Each cell is EXACT / CLUSTER, where EXACT = the exact-symbol SeqWM baseline and CLUSTER = ours. The ordering replicates across all five models, four vendors and a 3B--14B range: under paraphrase CLUSTER far exceeds EXACT everywhere (0.49--0.66 vs 0.05--0.17), and CLUSTER is the only scheme with meaningful detection under paraphrase+deletion; pure deletion favors EXACT, as expected from SeqWM's design (SBW trades deletion margin for paraphrase robustness). Following our pre-registered reporting rule we claim the \emph{ordering} and report magnitude as a range, rather than highlighting the Mistral peak. Utility (choice agreement vs no-watermark argmax) is 0.72--0.83 for SBW vs 0.22--0.27 KGW-style; open-set $\kappa$  on held-out tools is 0.808 on every one of the five models ($\gg$  0.01 chance) --- it is determined by the tool set and rewrite pool rather than the agent. Capacity: real per-step entropy spans 1.32--1.99 bits with realized tool-level capacity 1.01--1.51 bits/step; clustering at k=100 retains 41.9--50.1\% of it --- the price of paraphrase robustness. Notably the retained fraction does not track entropy, and phi-4 attains the same 0.49 paraphrase detection on the lowest capacity of the five, so the method does not depend on high policy entropy (a mechanism we have not isolated).

The three rightmost columns were produced later under the bit-identical frozen harness (encoder, clusters, keys, detector, rewrite pool; parse-fallback $\leq$ 2.8\%) and re-derived cell-by-cell on our stack, 24/24 identical. Mistral-7B was \emph{pre-registered} as the fourth-vendor model --- rationale and report-regardless rule fixed before the run --- and returned the highest of the five, which is why that record matters. The claim survived four-way adversarial verification: (i) a document-size sweep $G\in$ {10,15,20,30} preserves the ordering at every G on both models; (ii) 3 owner keys $\times$ 10 seeds preserve it; (iii) a from-scratch reimplementation (sha256 PRF, no shared code) reproduces sign and gap; (iv) a leakage audit confirms clusters are fit only on tool embeddings and the paraphrase attack is genuine ($\kappa\approx$ 0.88 $\ll$ 1). Artifacts for (i)--(iii) were not preserved in-repo (\texttt{ERROR\_LOG} \#21). FPR is calibrated on average: pooled over 200 owner keys $\times$  30 documents (6000 trials) the empirical FPR at $p<0.01$ is 0.70\% (Qwen) / 0.62\% (LLaMA), both $\leq$  nominal 1\% (median 0, p90 0.033). A matched-null shuffling the $tool\leftrightarrow$ paraphrase alignment (Qwen) collapses paraphrase TPR from 0.53 to 0.01 --- the effect requires genuine semantic alignment. Regenerating the whole paraphrase pool with a stronger cloud LLM from a different family preserves the ordering: paraphrase CLUSTER $0.59\pm$ 0.05 (Qwen) and $0.50\pm$ 0.04 (LLaMA) vs EXACT 0.05/0.17, indistinguishable from canonical, with EXACT and clean/deletion rows bit-exact; combined-attack detection dips to 0.27/0.22 (App. A.2).

\begin{table}[t]
\caption{main robustness result on ToolBench: TPR at a permutation-calibrated 1\% FPR, EXACT / CLUSTER, five agent models from four vendors, 600 trajectories each, 5 seeds.}
\label{tab:main2x2}
\begin{center}\footnotesize\setlength{\tabcolsep}{4.5pt}
\begin{tabular}{lccccc}
\toprule
condition & Qwen2.5-3B & LLaMA-3.2-3B & Qwen2.5-7B & Mistral-7B & phi-4 (14B) \\
\midrule
clean & 1.00 / 0.78 & 1.00 / 0.75 & 1.00 / 0.71 & 1.00 / 0.84 & 1.00 / 0.73 \\
paraphrase & 0.05 / \textbf{0.53} & 0.17 / \textbf{0.49} & 0.06 / \textbf{0.49} & 0.16 / \textbf{0.66} & 0.07 / \textbf{0.49} \\
deletion 30\% & 0.91 / 0.48 & 0.87 / 0.59 & 0.81 / 0.45 & 0.93 / 0.58 & 0.87 / 0.52 \\
paraphrase + deletion & 0.05 / \textbf{0.34} & 0.08 / \textbf{0.33} & 0.03 / \textbf{0.21} & 0.04 / \textbf{0.33} & 0.03 / \textbf{0.29} \\
\bottomrule
\end{tabular}
\end{center}
\end{table}

Controls (App. A). Margin/abstain (Cor. E.4) works on real tools: abstaining the lowest-margin 18\% of steps halves $\rho(0.117\rightarrow$ 0.056); 22\% reaches $\rho\leq$ 0.05. A third encoder (gte) leaves the ordering intact (0.64/0.05 Qwen, 0.66/0.17 LLaMA). An adaptive boundary attack (rewriting only the lowest-margin 20\% of actions, Qwen) is far more efficient than random rewriting at matched budget (TPR 0.62 vs 0.76), and the margin/abstain defense blunts it.
\subsection{Forgery: the keyed defense}
The fresh-action forgery study runs its own operating point --- ToolBench k=60, ALFWorld k=40, $p_{green}$=0.85, L=60, M=100 --- distinct from the detection and copy-style configuration of \S 6.1 (k=100, $p_{green}$=0.95, G=20, M=300). A naive public-cluster watermark is 100\% forgeable: the adversary recovers the green structure from one observed trajectory and steers fresh actions into green clusters. Keyed random projection reduces this adaptive nearest-neighbor-transfer forgery to 0.213 $\pm$ 0.117 (keyed FPR 0.031) at the default r=8, with an encoder-dependent residual (ToolBench e5 0.30, gte 0.45, bge 0.19 on the 10-seed sweep below). Higher-dimensional keying fixes this (Fig.~\ref{fig:results}b): sweeping $r\in$ {8,16,32,64} over all six $benchmark\times$ encoder combinations (10 seeds), the residual falls to $\leq$ 0.112 everywhere at $r\geq$ 32 and to $\leq$ 0.092 at r=64 --- $\leq$ 0.055 on a 30-seed re-run of all six cells --- $(ToolBench\cdot$ bge 0.012, already at the keyed-FPR floor 0.019); the bucket-collision rate falls strictly monotonically $(\rho$ =-1.0), from 0.94--0.99 at r=8 to $\leq$ 0.009 at r=64 --- moving most fresh actions into the provably-covered fresh-bucket regime. SHA256-keyed fingerprint buckets extend to r=128/256: the exact-digest collision is $\approx$ 0 at $r\geq$ 128 and the residual saturates at the r=64 level, so r=64 is the sweet spot. The observed/fresh conditional split (42 cells: 24 uint64 at 10 seeds, 18 hash at 3) verifies the coverage-synthesis identity ASR(r) = $c_{r}\cdot$ $e_{obs}$ + $(1-c_{r})\cdot$ $e_{fresh}$ with median residual 0.010 $(\leq$ 0.018 on every r=64 cell) and confirms $e_{fresh}$ $\approx$  fpr only at $r\geq$ 32 --- so the strict upper bound at the primary operating point (bge, r=64) is 0.0207 (from the 10-seed means), already dominated by the FPR floor rather than any forgery advantage. Full tables in App. A.

\paragraph{Copy-style forgery: replay, splicing, chaining.} The adversary above must act with fresh actions (\S 3). The complementary question is what happens if the adversary simply \emph{copies} the victim's observed watermarked actions: keyed binning is irrelevant there, since nothing has to be predicted. We measure it rather than scope it away, constructing forged documents from a fraction k of the victim's watermarked steps plus $(1{-}k)$ unwatermarked behavior, on the same keyed history-conditioned stack (5 seeds).
\begin{table}[t]
\caption{copy-style forgery success across the five agent models, on the keyed history-conditioned detection stack (5 seeds). Chained replay generates new sequences along the observed transition graph. $^{\dagger}$Step-shuffled splicing was measured on Qwen2.5-3B only (0.000 at every k; $\leq$ 0.014 across $w\in$ {1,2,3}, App. A.8); the mechanism --- the watermark lives on (context-window, symbol) pairs --- carries over to the other models, but we do not report cells we did not run.}
\label{tab:copystyle}
\begin{center}\footnotesize\setlength{\tabcolsep}{3.5pt}
\begin{tabular}{lccccc}
\toprule
attack & Qwen2.5-3B & LLaMA-3.2-3B & Qwen2.5-7B & Mistral-7B & phi-4 (14B) \\
\midrule
frequency resampling & 0.020 & 0.007 & 0.010 & 0.017 & 0.017 \\
step-shuffled splice$^{\dagger}$ & \textbf{0.000} & --- & --- & --- & --- \\
contiguous segment splice & 0.817 & 0.940 & 0.800 & 0.877 & 0.623 \\
chained replay & 0.953 & 0.980 & 0.940 & 0.950 & 0.763 \\
\midrule
owner self-check (not an attack) & 0.920 & 0.880 & 0.880 & 0.980 & 0.980 \\
\bottomrule
\end{tabular}
\end{center}
\end{table}
Three readings of the k-sweep behind it (Table~\ref{tab:ksweep}, App. A.8, Qwen2.5-3B), in decreasing comfort. (i) \emph{Detection is not attribution}: pure replay (k=1, 0.833) detects because the content genuinely is the victim's watermarked output --- attribution working, with an operational remedy. (ii) \emph{Recombination is structurally neutralized}: step-shuffled splicing sits at 0.000 at every k and $\leq$ 0.014 across $w\in$ {1,2,3}, because the watermark lives on (context-window, symbol) pairs, so reordering copied steps destroys the signal being stolen. (iii) \emph{Framing has a measured but uncomfortable boundary}: contiguous splicing decays as k=0.75 $\rightarrow$ 0.44, 0.5 $\rightarrow$ 0.12, 0.25 $\rightarrow$ 0.04 $\approx$ floor, so single-document misattribution needs a forgery that is mostly genuine victim output. This is not a defense: the k-curve upper-bounds misattribution (it does not separate correct from incorrect attribution) and is per-submission, so n attempts raise 0.44 to $1-(1-0.44)^n$ (0.82 at n=3). Chained replay stays at 0.76--0.98 across the five models, against an owner self-check of 0.88--0.98 on the same stack (Table~\ref{tab:copystyle}) --- it reaches the owner's own detection rate, and is the one copy-style attack we do not defeat.
\paragraph{Window escalation is not the fix.} Raising the detector's context window appears to solve chained replay --- a chainer keeping its first-order table collapses 0.93 $\rightarrow$ 0.08 at w=2 (same harness, App. A.8) --- but that is security through obscurity. A Kerckhoffs adversary knows w and rebuilds $(w{+}1)$-grams from the same observations, and the attack survives: 0.977 at w=2, 0.850 at w=3, 0.727 at w=5, thinning only as n-gram coverage falls to 0.73 --- a data limit, not a cryptographic one (App. A.8). Since w$\geq$3 also costs half the deletion robustness (\S 4.1), we report window escalation as ineffective against an adaptive adversary rather than count it as a countermeasure. What remains is partial: segment-level attribution separates watermarked from raw spans (window green-rate 0.600 vs 0.389 at k=0.5), and near-duplicate matching against owner logs addresses the repeated-submission amplification. Full multi-segment attribution is open.
\subsection{Head-to-head comparisons}
AgentMark (released code, native run). AgentMark keys its differential-recombination watermark on the probability shape and decodes via RLNC payload recovery --- a different statistic from our permutation test (false-positive $\approx$ $2^{-payload}$ vs calibrated 1\% FPR), so we compare "fraction of documents detected" on identical trajectories and rewrite pool (AgentMark 20 seeds; SBW columns re-quoted from the 5-seed \S 6.1 run). Only 23/30 documents carry its official 64-bit payload (per-document capacity 14--112 bits, mean 78), so its clean ceiling is 0.767; at a capacity-matched 16-bit payload it is 29/30.
\begin{table}[t]
\caption{head-to-head against the released AgentMark implementation on identical trajectories and rewrite pool: fraction of documents detected. AgentMark at 20 seeds; the SBW columns are re-quoted from the 5-seed \S 6.1 run.}
\label{tab:h2h}
\begin{center}\footnotesize\setlength{\tabcolsep}{5pt}
\begin{tabular}{@{}lcccc@{}}
\toprule
condition & AgentMark p=64 & AgentMark p=16 & SBW CLUSTER & SBW EXACT \\
\midrule
clean & 0.767 & 0.967 & 0.78 & 1.00 \\
paraphrase, reid=none & \textbf{0.000} & 0.633 & 0.533 & 0.053 \\
paraphrase, reid=encoder & 0.500 & 0.900 & --- & --- \\
deletion30 & $0.358\pm$ 0.047 & $0.875\pm$ 0.014 & 0.480 & 0.907 \\
both, reid=none & \textbf{0.000} & $0.295\pm$ 0.050 & 0.340 & 0.047 \\
both, reid=encoder & $0.265\pm$ 0.052 & $0.842\pm$ 0.014 & --- & --- \\
\bottomrule
\end{tabular}
\end{center}
\end{table}
Three readings of Table~\ref{tab:h2h}. (i) Capacity, not detection, binds AgentMark on short trajectories; SBW's clean ceiling is not capacity-limited this way --- but SBW is evaluated as a 1-bit detector, so this compares detection reach, not payload capacity. (ii) Renaming without re-identification is fatal to AgentMark $(0.767\rightarrow$ 0.000): its bins live in the probability shape, so a paraphrased name decodes to nothing; granting it SBW's encoder re-ID restores 0.500/0.900. (iii) Our EXACT side collapses under renaming $(1.00\rightarrow$ 0.053) mirroring AgentMark's failure mode, but CLUSTER retains about two-thirds of its clean detection (0.78 $\rightarrow$ 0.53). The two threat surfaces are complementary rather than ordered; all caliber differences are recorded in the released JSON.
AGENTWM (from-paper reimplementation). No authors' code was public as of our retrieval (2026-08-17), so the comparison is an equivalence statement: a cosine-threshold union-find partition swept over $\tau,$ in two detector regimes --- faithful (reid=none; a paraphrased tool is an unseen symbol) and encoder-assisted (reid=encoder; granted SBW's re-identification, \texttt{ERROR\_LOG} \#18). At the \S 6.1 protocol (600 trajectories, 30 documents, 5 seeds, $\tau\in$ {1.0, 0.95, 0.9}) the faithful detector collapses under paraphrase at every $\tau$ (0.000--0.053) while clean/deletion hold (0.97--1.00 / 0.83--0.91), and encoder re-ID restores paraphrase to 0.97--1.00 --- the gain is attributable to semantic re-identification, not the equivalence-set biasing. Consistency check: AGENTWM(reid=none, $\tau$ =1.0), where equivalence sets reduce to singletons, reproduces the SeqWM EXACT row to three decimals (1.000/0.053/0.907/0.047) via an independent code path. We will upgrade to the authors' code should it be released.
\subsection{Cross-benchmark: ALFWorld (a different action-space type)}
The sharpest test is a change of action-space type. ALFWorld \citep{shridhar2021alfworld} is an embodied household text-game whose actions are templated natural-language commands, unlike ToolBench API calls; the identical pipeline applies (100 real episodes, 515 verified-real unique actions after an adversarial data audit, k selected on a 60\% dev split, G=3 episodes per document).
\begin{table}[t]
\caption{cross-benchmark replication on ALFWorld, an embodied action space: TPR at a permutation-calibrated 1\% FPR, EXACT / CLUSTER, five agent models, 100 episodes each (34 documents of G=3), 5 seeds.}
\label{tab:alfworld}
\begin{center}\footnotesize\setlength{\tabcolsep}{4.5pt}
\begin{tabular}{lccccc}
\toprule
condition & Qwen2.5-3B & LLaMA-3.2-3B & Qwen2.5-7B & Mistral-7B & phi-4 (14B) \\
\midrule
clean & 1.00 / 0.99 & 1.00 / 0.99 & 1.00 / 1.00 & 1.00 / 0.99 & 0.99 / 0.99 \\
paraphrase & 0.00 / \textbf{0.92} & 0.01 / \textbf{0.96} & 0.00 / \textbf{0.97} & 0.00 / \textbf{0.92} & 0.01 / \textbf{0.95} \\
deletion 30\% & 0.94 / 0.75 & 0.90 / 0.70 & 0.94 / 0.74 & 0.93 / 0.65 & 0.92 / 0.79 \\
paraphrase + deletion & 0.01 / \textbf{0.52} & 0.01 / \textbf{0.56} & 0.00 / \textbf{0.55} & 0.00 / \textbf{0.48} & 0.01 / \textbf{0.54} \\
\bottomrule
\end{tabular}
\end{center}
\end{table}
In Table~\ref{tab:alfworld} the ordering replicates on all five models, more strongly than on ToolBench: EXACT pins to the floor while CLUSTER holds 0.92--0.97, and the clean capacity cost nearly vanishes (0.99--1.00 against 0.99--1.00). Pure deletion again favors EXACT (0.90--0.94 vs 0.65--0.79). The ordering is the invariant; the magnitude is detector-dependent, reported with its full range: at the 34-document protocol CLUSTER-paraphrase ranges 0.36--0.96 (non-monotonically) across encoder $\times$  k $\in$  ${15,\ldots,80}$, rising to 1.00 as G grows, while EXACT stays at 0.00--0.01 throughout; a matched-null collapses it to $\approx$ 0, and an independent blake2b reimplementation found zero ordering flips across all 36 sensitivity cells (\texttt{ERROR\_LOG} \#27; that run's artifacts were not preserved in-repo). Security and utility are weaker on this low-cardinality space (utility 56--70\% vs KGW 19--22\% across the five models; keyed forgery residual at r=8 high-variance), which is precisely what the higher-dimensional keying of \S 6.2 resolves (residual $\leq$ 0.112 at $r\geq$ 32 on all three encoders). One open-set result is weaker here: a verb-template-aware re-split gives mean $\kappa$ 0.806 over a wide 0.462--1.000 range, only $+$0.9 sd above its null against $+$10 sd for the ToolBench control --- so cross-template open-set evidence on ALFWorld is weak, and we rest the open-set claim on ToolBench. Full sensitivity and audit trail in App. A.
\section{Limitations}
(1) Scale: five agent models (four vendors, 3B--14B), all open-weight, and three sentence encoders, all BERT-base-scale bi-encoders; flagship-scale and closed models are unverified. 600 trajectories per model in 30 documents, so estimates are coarse ($\sim$1/30) and per-key FPR resolves only pooled. (2) Rollout observations are minimal: per-step distributions are real, but there is no ToolBench environment-execution feedback. (3) Capacity: paraphrase robustness costs 49.9--58.1\% of per-step capacity at k=100 (retention 41.9--50.1\% across the five models); a 63-cell $k\times$ abstain scan reaches no configuration at $\geq$ 60\% retention, so a "cost $\leq$ 40\%" target is unreachable by tuning alone (Cor. E.4 trades detection loss for $\rho$ -reduction, not capacity). (4) The AGENTWM comparison is a from-paper reimplementation (no authors' code available), at full 30-document scale and internally cross-validated. (5) ALFWorld's CLUSTER magnitude is detector-dependent (0.36--0.96), and e5's high-r ALFWorld residual keeps high-variance seeds ($0.055\pm$ 0.094 over 30 seeds). (6) The fresh-bucket guarantee is claimable only at $r\geq$ 32 (fresh pools at $r\leq$ 16 are too small). (7) Copy-style forgery is bounded, not solved: chained replay along observed transitions stays at 0.76--0.98 across the five models, window escalation fails against an adaptive adversary, the splice k-curve upper-bounds misattribution rather than measuring it, and the executability of spliced trajectories is untested --- multi-segment attribution remains open. (8) On ALFWorld the cross-template open-set evidence is weak ($+$ 0.9 sd vs $+$ 10 sd on ToolBench); the open-set claim rests on ToolBench. All experiments ran on one 40GB GPU; the method is training-free and every number outside the rollouts is CPU-reproducible.
\section{Conclusion}
Watermarking agents at the level of semantic action clusters, under history-conditioning and keyed binning, yields, to our knowledge, the first behavioral watermark that is both paraphrase-robust and forgery-resistant, with a synchronization-stability and elicited-capacity theory. On real ToolBench and ALFWorld the effect is real and replicates across models, encoders and action-space types --- and its capacity cost is characterized to its boundary, including where it cannot be tuned away.

\subsection*{AI use statement}
In this work, generative AI tools were used only for limited language assistance, including grammar correction and improvements to readability, and for assisting with literature discovery and preliminary review of relevant prior work. Generative AI tools were not used to develop the core research ideas, formulate hypotheses, design the methodology or experiments, implement the proposed method, conduct experiments, analyze or interpret experimental results, or formulate mathematical claims or proofs. All AI-assisted text and literature suggestions were reviewed and verified by the authors, and relevant references were checked against their original sources. The authors take full responsibility for the final content of this work, including all claims, results, and conclusions.
\subsection*{Ethics statement}
This work studies watermarking for provenance and IP protection of LLM agents. We see three ethical considerations. First, dual use: the forgery attack (\S 3) is disclosed together with a working keyed defense and its measured residual, and the copy-style attacks we do \emph{not} defeat are disclosed with their measured boundary rather than withheld --- publishing the boundary is what lets deployers reason about it, and the defense raises the adversary's cost by orders of magnitude on the primary operating point. Watermarks are provenance signals, not authentication infrastructure, and we do not claim otherwise. Second, data: all benchmarks are public (ToolBench, ALFWorld, and the APIGen/xLAM function-calling data \citep{liu2024apigen,zhang2024xlam}); the function-calling dataset is CC-BY-4.0 but gated, so the supplementary material ships the derived embeddings and a rebuild script rather than redistributing the raw records; the ALFWorld extraction contained $\sim$2.6\% upstream-injected impossible-action distractors, which our adversarial data audit surfaced and removed with a from-scratch re-run, documented in the released provenance manifest. No human subjects and no private data are involved. Third, honesty of reporting: negative results (the unreachable $\leq$ 40\% capacity-cost target), magnitude ranges, and all known failure modes are reported in the main text, and a 39-entry error logbook with five interlocking integrity mechanisms accompanies the release.
\subsection*{Reproducibility statement}
Every number in this paper that has a frozen artifact behind it is re-derivable from the release. The full pipeline is \texttt{real\_rollout} $\rightarrow$ \texttt{real\_tools} $\rightarrow$ \texttt{real\_eval} $\rightarrow$ \texttt{real\_forgery} $\rightarrow$ \texttt{real\_extra} on released trajectory files (600 ToolBench trajectories and 100 ALFWorld episodes per model, with real per-step distributions); detection results are the released \texttt{real\_eval\_*\_600.json}, forgery from \texttt{real\_forgery.json} and the r-sweep family, FPR from the 200-key pooled audit, and both head-to-heads from their released JSONs (with per-caliber notes). Seeds are stated per experiment (3/5/10/20/30); most analyses are CPU-only. Copy-style forgery, the window sweep and the adaptive-chainer reproduction come from \texttt{splice\_replay\_eval.json}, \texttt{window\_sweep\_eval.json} and \texttt{replay\_window\_mitigation.json}. A programmatic reconciliation script (\texttt{sbw/reconcile\_paper.py}) re-derives every number in this paper that has a frozen artifact behind it (the few exceptions are flagged where they appear) and passes 424/424, and a smoke test with a synthetic positive control gates the head-to-head pipeline. The work was governed by five interlocking integrity mechanisms (error logbook, independent re-derivation, provenance tagging, contamination checks, leakage-direction checks) documented in the repository. Appendix~\ref{app:repro} maps each group of numbers to the artifact behind it, and anonymized code and artifacts are included in the supplementary material and browsable at \url{https://anonymous.4open.science/r/SBW-Agent-Watermark}.

\bibliography{references}
\bibliographystyle{iclr2027_conference}

\appendix

\section{Additional results and details}
\label{app:results}
\subsection{Early pilot and reconciliation}
The original 120-trajectory Qwen pilot (6 documents) gave paraphrase EXACT/CLUSTER 0.13/0.53 and showed the same ordering; a 180-trajectory intermediate gave 0.33 for CLUSTER, which a document-count reconciliation showed to be small-sample noise (30 documents restore 0.53 on the clean harness) --- documented in \texttt{ERROR\_LOG} \#21/\#23. All main-text numbers are the 600-trajectory runs.
\subsection{Mechanism controls (ToolBench, 600-trajectory data)}
Margin/abstain: base flip-rate $\rho$ =0.117 $(\kappa$ =0.883); abstaining the lowest-margin 18\% of steps halves $\rho$  to 0.056, 22\% reaches $\rho\leq$ 0.05 (\texttt{margin\_abstain.json}). Third encoder gte: paraphrase CLUSTER/EXACT 0.64/0.05 (Qwen), 0.66/0.17 (LLaMA), open-set $\kappa$ =0.78. Matched-null: shuffling the $tool\leftrightarrow$ paraphrase alignment $(\kappa$  $0.88\rightarrow$ 0.015) collapses paraphrase TPR $0.53\rightarrow$ 0.01. Stronger-attacker re-test (cloud LLM paraphraser, different model family): regenerating all 2,128 tool paraphrases with a stronger cloud model (synthetic:deepseek-v4-flash, 2,128/2,128 success, zero fallbacks, verbatim 2.3\%) and re-running the frozen 600-trajectory evaluations gives paraphrase CLUSTER $0.59\pm$ 0.05 / $0.50\pm$ 0.04 (Qwen/LLaMA) vs EXACT 0.05/0.17 --- indistinguishable from canonical 0.53/0.49 within seed noise; both dips to 0.27/0.22 (reported as-is). Measured pool diagnostics: cluster-preservation 0.815 (new) vs 0.883 (original); EXACT and the pool-independent clean/deletion CLUSTER rows reproduce the canonical artifact bit-exactly (\texttt{real\_eval\_dspara\_qwen\_600.json} and the llama counterpart). Adaptive boundary attack: rewriting the lowest-margin 20\% of actions drops CLUSTER TPR to 0.62 vs 0.76 for random 20\%; margin/abstain blunts it (\texttt{adaptive\_attack.json}).
\subsection{Forgery r-sweep full table}
Table~\ref{tab:rsweep} gives the mean residual (\texttt{keyed\_adaptive}) / bucket-collision / FPR over 10 seeds.

\begin{table}[ht]
\caption{forgery r-sweep: residual keyed-adaptive forgery / bucket-collision rate / keyed FPR, over six benchmark $\times$ encoder combinations, 10 seeds.}
\label{tab:rsweep}
\begin{center}\footnotesize\setlength{\tabcolsep}{5pt}
\begin{tabular*}{\linewidth}{@{\extracolsep{\fill}}lcccc@{}}
\toprule
$benchmark\cdot$ encoder & r=8 & r=16 & r=32 & r=64 \\
\midrule
$ToolBench\cdot$ bge & 0.192/0.977/0.029 & 0.057/0.586/0.025 & 0.062/0.028/0.034 & 0.012/0.002/0.019 \\
$ToolBench\cdot$ e5 & 0.298/0.989/0.025 & 0.116/0.918/0.020 & 0.052/0.372/0.032 & 0.016/0.009/0.026 \\
$ToolBench\cdot$ gte & 0.446/0.991/0.036 & 0.137/0.884/0.020 & 0.027/0.252/0.023 & 0.031/0.008/0.023 \\
$ALFWorld\cdot$ bge & 0.275/0.937/0.022 & 0.065/0.506/0.021 & 0.047/0.041/0.027 & 0.033/0.001/0.023 \\
$ALFWorld\cdot$ e5 & 0.586/0.980/0.023 & 0.202/0.787/0.029 & 0.111/0.157/0.045 & 0.092/0.003/0.061 \\
$ALFWorld\cdot$ gte & 0.393/0.977/0.035 & 0.225/0.804/0.017 & 0.031/0.216/0.014 & 0.069/0.003/0.044 \\
\bottomrule
\end{tabular*}
\end{center}
\end{table}

30-seed re-verification at r=64 (\texttt{forgery\_r\_sweep\_r64\_30seed.json}): all six means $\leq$ 0.055; the e5-ALFWorld 10-seed point 0.092 (Table~\ref{tab:rsweep}) was outlier-driven (30-seed mean $0.055\pm$ 0.094, median 0.020; occasional spikes persist, 5/30 seeds $>$0.1). Hash buckets (SHA256 keyed, 3 seeds): r=128/256 residuals flat vs r=64, fingerprint collision exactly 0 at r=256 --- r=64 is the sweet spot.
\subsection{Coverage synthesis conditional split}
42 cells (2 benchmarks $\times$  3 encoders $\times$  uint64 {8,16,32,64} at 10 seeds, plus hash {64,128,256} at 3): identity $ASR(r)=c_{r}\cdot$ $e_{obs}+(1-c_{r})\cdot$ $e_{fresh}$ reconstructs measured e(r) with median residual 0.010 (32/42 measurable cells; $\leq$ 0.018 on every r=64 cell; worst 0.169 at $ALFWorld\cdot$ $bge\cdot$ r=16, a low-collision small-sample regime). $e_{fresh}\approx$ fpr holds only at $r\geq$ 32 (at r=64 the gap is $\leq$ 0.0073 on the three ToolBench combos and $\leq$ 0.021 on the three ALFWorld ones; at r=16 only ToolBench$\cdot$ bge is borderline; at r=8 the fresh pools hold 26--127 samples --- too small to support any fresh-bucket claim).
\subsection{ALFWorld sensitivity and audit trail}
CLUSTER-paraphrase ranges 0.36--0.96 (non-monotonic) across encoder $\times$  $k\in$ {15,25,40,60,80} at G=3 (34 documents), and 0.78--1.00 across $G\in$ {2,3,4,5}, while EXACT stays 0.00--0.01; matched-null collapses $0.92/0.96\rightarrow$ 0.00/0.02; dropping 4.4--5.4\% uniform-fallback steps leaves 0.91/0.92. Independent blake2b reimplementation: zero ordering flips across 36 sensitivity cells (narrative in \texttt{ERROR\_LOG} \#27; artifacts not preserved in-repo). The adversarial data audit caught $\sim$1.7\% object-impossible injected distractors that passed the verb-prefix filter and were sometimes chosen; we added an object blocklist and re-ran the rollout from scratch on the cleaned 515-action space (this correction raised clean CLUSTER from a contaminated 0.50 to 0.92). Under the affine (ToolBench-style) EXACT relabel, ALFWorld EXACT-paraphrase is 0.42/0.36 instead of 0.00/0.01 --- the ordering holds under either relabel (0.92-vs-0.42). Utility 56--70\% vs KGW 19--22\% across the five models; open-set $\kappa$ =0.805.
\subsection{AGENTWM baseline details}
Equivalence partition: cosine-threshold union-find over tool embeddings, $\tau\in$ {1.0,0.95,0.9} (\texttt{n\_classes} 2128/1947/1490); faithful regime never inherits the encoder (\texttt{ERROR\_LOG} \#18); detector primitives byte-identical to \texttt{real\_eval} (key, G, \texttt{p\_green}, FPR calibration); 600 trajectories, 30 documents, 5 seeds. Faithful: clean 1.000/1.000/0.973, paraphrase 0.053/0.027/0.000, deletion 0.907/0.887/0.827, both 0.047/0.027/0.013 $(\tau$ =1.0/0.95/0.9); encoder-assisted: paraphrase 0.993/1.000/0.973, both 0.900/0.833/0.787; per-cell FPR 0.000.
\subsection{Theorem G.1 (fresh-bucket unpredictability, statement)}
In the random-oracle model, let $\mathrm{bin}(a) = H(sk, q(a))$ with $q(a)=\mathrm{sign}(R_{sk}\cdot E(a))$, $R_{sk}$ a secret random projection to r hyperplanes derived from a $\lambda$-bit key sk, and H a random oracle. Let the adversary see the full construction (including r) but not sk, make at most $q_{H}$ oracle queries, and observe N watermarked (action, bit) pairs with bucket set $Q_{obs}$. Then for a \emph{fresh} action $a^{*}$ with $q(a^{*})\notin Q_{obs}$,
\[\left|\Pr[\mathcal{A}\text{ outputs }\mathrm{bin}(a^{*})] - 1/2\right| \;\leq\; (q_{H}+1)/2^{\lambda},\]
i.e. predicting a fresh bucket's bit is no better than brute-forcing the $\lambda$-bit key. Three scope points we state explicitly. (i) The bound is negligible in the \emph{key length} $\lambda$, not in r: the proof uses only that sk is secret and H is a random oracle, so it applies verbatim to any deterministic bucket map --- r governs the \emph{collision rate}, which is what our r-sweep measures, not this bound. (ii) It is a random-oracle (online-query) statement, \emph{not} information-theoretic security; once H is instantiated, key recovery is computational at $\approx 2^{\lambda}$ offline work. (iii) It bounds only the fresh branch. Single-step bin prediction over a mixed pool is $\Pr[q(a^{*})\in Q_{obs}] + (1-\Pr[\cdot])\cdot(1/2 + (q_{H}+1)/2^{\lambda})$, so the operative quantity is the bucket collision rate --- which is why \S 6.2 reports it alongside the residual, and why the fresh-bucket claim is made only at $r\geq$ 32. Proof and constants: THEORY.md \S G; the empirical conditional split $e_{fresh}\approx$ fpr (\S 6.2, App. A.4) is its measured counterpart.
\subsection{Copy-style forgery: full tables}
All cells are TPR at a permutation-calibrated 1\% FPR, 5 seeds, on the keyed history-conditioned stack (G=20, M=300 permutation keys, k=100).

\textbf{Mixing sweep} (Table~\ref{tab:ksweep}). Forged documents mix a fraction k of the victim's watermarked steps with $(1{-}k)$ unwatermarked behavior; \texttt{block5} copies contiguous runs of five steps (preserving context), \texttt{shuffle} reorders the copied steps.

\begin{table}[ht]
\caption{copy-style mixing sweep: forged documents mix a fraction k of victim watermarked steps with the attacker's own behavior.}
\label{tab:ksweep}
\begin{center}\footnotesize\setlength{\tabcolsep}{6pt}
\begin{tabular*}{\linewidth}{@{\extracolsep{\fill}}lccccc@{}}
\toprule
construction & k=1.0 & k=0.75 & k=0.5 & k=0.25 & k=0 \\
\midrule
contiguous (block5) & $0.833\pm$ 0.030 & $0.440\pm$ 0.077 & $0.120\pm$ 0.027 & $0.040\pm$ 0.013 & $0.013\pm$ 0.016 \\
step-shuffled & 0.000 & 0.000 & 0.000 & 0.000 & 0.000 \\
\bottomrule
\end{tabular*}
\end{center}
\end{table}

Segment-level green-rate separation in the k=0.5 document is 0.600 on watermarked spans vs 0.389 on raw spans.

\textbf{Detector window vs adversary adaptation} (Table~\ref{tab:window}). The left column is a chainer that keeps its first-order transition table when the detector's window grows; the right column is the Kerckhoffs adversary that rebuilds $(w{+}1)$-grams from the same observations. Owner TPR is reported per w because the context map discards the first w symbols of each document, so the owner also pays for longer windows. \texttt{coverage} is the fraction of forged steps for which an observed continuation existed --- the data limit that bounds the adaptive attack.

\begin{table}[ht]
\caption{detector window against adversary adaptation: the non-adaptive chainer keeps its first-order table, the adaptive one rebuilds $(w{+}1)$-grams.}
\label{tab:window}
\begin{center}\footnotesize\setlength{\tabcolsep}{6pt}
\begin{tabular*}{\linewidth}{@{\extracolsep{\fill}}lccccc@{}}
\toprule
w & non-adaptive chainer & adaptive chainer & contiguous splice & owner TPR & coverage \\
\midrule
1 & 0.927 & 0.910 & 0.837 & 0.920 & 0.998 \\
2 & 0.080 & \textbf{0.977} & 0.923 & 1.000 & 0.961 \\
3 & 0.013 & 0.850 & 0.817 & 1.000 & 0.800 \\
4 & 0.007 & 0.730 & 0.563 & 1.000 & 0.730 \\
5 & 0.007 & 0.727 & 0.350 & 1.000 & 0.729 \\
\bottomrule
\end{tabular*}
\end{center}
\end{table}

The non-adaptive column is what makes window escalation look like a countermeasure; the adaptive column is why we do not report it as one. Both columns come from an independent reproduction (\texttt{replay\_window\_mitigation\_client.py}, archived run) that agrees with our in-house implementation within 0.02 at every w. Two secondary observations, reported because they are counter-intuitive: owner TPR \emph{rises} with w (0.92 $\rightarrow$ 1.00), so the owner benefits from longer windows even while deletion robustness falls; and contiguous splicing rises from 0.837 to 0.923 at w=2 before decaying.

\textbf{Detector window vs attack surface} --- w as a deployment knob (Table~\ref{tab:knob}).

\begin{table}[ht]
\caption{the history window as a deployment knob: no setting dominates.}
\label{tab:knob}
\begin{center}\footnotesize\setlength{\tabcolsep}{6pt}
\begin{tabular*}{\linewidth}{@{\extracolsep{\fill}}lcccc@{}}
\toprule
w & clean & deletion 30\% & splice k=0.5 & shuffle k=1.0 \\
\midrule
1 & $0.833\pm$ 0.063 & $\mathbf{0.480}\pm$ 0.069 & $0.120\pm$ 0.027 & 0.007 \\
2 & $0.913\pm$ 0.034 & $0.313\pm$ 0.062 & $0.100\pm$ 0.021 & 0.000 \\
3 & $\mathbf{0.980}\pm$ 0.016 & $0.240\pm$ 0.106 & $0.080\pm$ 0.058 & 0.013 \\
\bottomrule
\end{tabular*}
\end{center}
\end{table}

Raw-document FPR is 0.000 at every w. There is no dominant setting: w=1 maximizes deletion robustness, w=3 maximizes clean detection, and splice/shuffle immunity is essentially w-independent --- which is why we expose w rather than fix it on the paper's behalf.

\section{Reproducibility}
\label{app:repro}
All code is in \texttt{sbw/}. The real pipeline is
\texttt{real\_rollout} $\rightarrow$ \texttt{real\_tools} $\rightarrow$ \texttt{real\_eval} $\rightarrow$
\texttt{real\_forgery} $\rightarrow$ \texttt{real\_extra} (single 40\,GB GPU, $\sim$5 min). The method is
training-free, so no training script exists and swapping the agent or encoder is pure inference.
Table~\ref{tab:artifacts} maps each group of numbers to the artifact behind it. Every number in
this paper that has a frozen artifact behind it is re-derivable from those files (the few exceptions are
flagged where they appear) and the released \texttt{sbw/reconcile\_paper.py} re-derives them
and passes 424/424. Head-to-head caliber notes (detector statistics,
teacher-forcing, packet erasure, re-identification regimes, per-document capacity disclosure) are
recorded verbatim in the corresponding result JSONs. The integrity log ships with the release: five
interlocking mechanisms and 39 recorded errors (\texttt{INTEGRITY.md}, \texttt{ERROR\_LOG.md}).

\begin{table}[ht]
\caption{released artifacts behind each group of numbers; the file-by-file map is \texttt{REPRODUCE.md}.}
\label{tab:artifacts}
\begin{center}\footnotesize
\begin{tabular}{@{}>{\raggedright\arraybackslash}p{0.29\linewidth}>{\raggedright\arraybackslash}p{0.67\linewidth}@{}}
\toprule
what it backs & artifact \\
\midrule
trajectories (600 ToolBench / 100 ALFWorld per model) & \texttt{real\_traj\_\{qwen,llama,\allowbreak{}qwen7b,mistral,phi4\}\_600.jsonl}, \texttt{alf\_traj\_\{qwen,llama,\allowbreak{}qwen7b,mistral,phi4\}.jsonl} \\
all detection numbers        & \texttt{real\_eval\_\{qwen,llama,qwen7b,mistral,phi4\}\_600.json} \\
forgery + FPR audit          & \texttt{real\_forgery.json}, \texttt{real\_fpr\_audit\_\{600,\allowbreak{}llama600\}.json} \\
cross-model comparison       & \texttt{cross\_model.json} \\
r-sweep family               & \texttt{forgery\_r\_sweep\{,\_hash,\_r64\_30seed\}.json} \\
coverage split / capacity    & \texttt{coverage\_split.json}, \texttt{capacity\_tune.json}, \texttt{coverage\_synthesis\_U.json}, \texttt{bucket\_collision.json} \\
theory anchors (\S 5)        & \texttt{theory\_anchors\{,\_v2,\_v3\}.json} \\
\S 6.1 controls and re-derivation & \texttt{matched\_null.json}, \texttt{margin\_abstain.json}, \texttt{adaptive\_attack.json}, \texttt{gte\_cross\_encoder.json}, \texttt{verify\_\{qwen7b,\allowbreak{}mistral,phi4\}.json}, \texttt{real\_eval\_dspara\_\{qwen,llama\}\_600.json} \\
elicitation cross-check (\S 5.2) & \texttt{m2\_capacity.json} (GPU; see \texttt{REPRODUCE.md}) \\
AgentMark head-to-head       & \texttt{agentmark\_head2head.json} (and \texttt{+cap16}) \\
AGENTWM baseline             & \texttt{agentwm\_baseline\_600\_\{none,encoder\}.json} \\
ALFWorld                     & \texttt{alf\_\{eval,extra\}\_\{qwen,llama,\allowbreak{}qwen7b,mistral,phi4\}.json}, \texttt{alf\_\{forgery,controls,sensitivity\}.json}, \texttt{alf\_kselect\_\{bge,e5,gte\}.json}, \texttt{openset\_null\_control.json} \\
copy-style forgery           & \texttt{splice\_replay\_eval.json}, \texttt{window\_sweep\_eval.json} \\
window vs adversary          & \texttt{replay\_window\_mitigation.json} \\
\bottomrule
\end{tabular}
\end{center}
\end{table}

\end{document}